\documentclass[
 aps,
 11pt,
 final,
 notitlepage,
 oneside,
 onecolumn
 nobibnotes,
 nofootinbib,
 superscriptaddress,
 noshowpacs,
 centertags]
{revtex4}

\def\saoname{Special Astrophysical Observatory,  Russian Academy of Sciences,
              Nizhnii Arkhyz, 369167 Russia}

\def\squareforqed{\hbox{\rlap{$\sqcap$}$\sqcup$}}

\def\sq{\ifmmode\squareforqed\else{\unskip\nobreak\hfil
\penalty50\hskip1em\null\nobreak\hfil\squareforqed
\parfillskip=0pt\finalhyphendemerits=0\endgraf}\fi}

\def\degr{\hbox{$^\circ$}}

\def\arcmin{\hbox{$^\prime$}}

\def\arcsec{\hbox{$^{\prime\prime}$}}

\def\utw{\smash{\rlap{\lower5pt\hbox{$\sim$}}}}

\def\udtw{\smash{\rlap{\lower6pt\hbox{$\approx$}}}}

\def\fm{\hbox{$\,.\!\!^{\rm m}$}}

\def\fdg{\hbox{$\,.\!\!^\circ$}}

\def\farcm{\hbox{$\,.\mkern-4mu^\prime$}}

\def\farcs{\hbox{$\,.\!\!^{\prime\prime}$}}

\def\diameter{{\ifmmode\mathchoice
{\ooalign{\hfil\hbox{$\displaystyle/$}\hfil\crcr
{\hbox{$\displaystyle\mathchar"20D$}}}}
{\ooalign{\hfil\hbox{$\textstyle/$}\hfil\crcr
{\hbox{$\textstyle\mathchar"20D$}}}}
{\ooalign{\hfil\hbox{$\scriptstyle/$}\hfil\crcr
{\hbox{$\scriptstyle\mathchar"20D$}}}}
{\ooalign{\hfil\hbox{$\scriptscriptstyle/$}\hfil\crcr
{\hbox{$\scriptscriptstyle\mathchar"20D$}}}}
\else{\ooalign{\hfil/\hfil\crcr\mathhexbox20D}}%
\fi}}

\newcommand{\bsao}{Bull. Spec. Astrophys. Obs. }
\newcommand{\pasp}{Publ. Astron. Soc. Pacific }
\newcommand{\alet}{Astronomy Letters }
\begin{document}
\selectlanguage{english}

\title{\huge SCORPIO-2 Guiding and Calibration System in  the Prime Focus of the 6-m Telescope}

\author{\firstname{V.~L.}~\surname{Afanasiev}}
 \email{vafan@sao.ru}
 \affiliation{\saoname}

\author{\firstname{V.~R.}~\surname{Amirkhanyan}}
\affiliation{\saoname} \affiliation{Sternberg Astronomical Institute,
Lomonosov Moscow State University, Moscow, 119991 Russia}

\author{\firstname{A.~V.}~\surname{Moiseev}}
 \affiliation{\saoname}

\author{\firstname{R.~I.}~\surname{Uklein}}
 \affiliation{\saoname}

\author{\firstname{A.~E.}~\surname{Perepelitsyn}}
 \affiliation{\saoname}

\received{July 10, 2017} \revised{September 14, 2017}

\begin{abstract}
We describe a device (adapter) for off-axis guiding and photometric
calibration of wide-angle spectrographs operating in the prime focus
of the 6-m telescope of the Special Astrophysical Observatory of the
Russian Academy of Sciences. To compensate coma in off-axis star
images an achromatic lens corrector is used, which ensures
maintaining image quality  ($FWHM$) at a level of about $1\arcsec$
within $15\arcmin$ from the optical axis. The device has two
$54\arcsec$-diameter movable guiding fields, which can move in
$10\arcmin\times4\farcm5$ rectangular areas. The device can perform
automatic search for guiding stars, use them to  control the
variations of atmospheric transmittance, and focus the telescope
during exposure. The limiting magnitude of potential guiding stars is
$m_R\sim17^{\mathrm m}$. The calibration path whose optical
arrangement meets the telecentrism condition allows the spectrograph
to be illuminated both by a source of line spectrum (a He--Ne--Ar
filled lamp) and by a source of continuum spectrum. The latter is
usually represented either by a halogen lamp or a set of
light-emitting diodes, which provide illumination of approximately
uniform intensity over the wavelength interval from 350 to 900~nm.
The adapter is  used for observations with  SCORPIO-2 multimode focal
reducer.
\end{abstract}

\maketitle

\section{INTRODUCTION}

The six-meter Big Telescope Alt-Azimuth (BTA) is the first large
telescope on an altazimuth mounting with machine-controlled tracking
of target objects~\cite{BTA71}. This  was a pioneering technical
solution for up to the time (late 1960s) which gave rise to the use
of altazimuth mounts for large, massive telescopes. Besides the
specifics of guiding, which required the use of a computer-assisted
control system, altazimuth mounting is, unlike parallactic mounting,
distinguished by the rotation of the field of view. Note that the
accuracy of the tracking of the observing object is determined not
only by the quality of telescope mechanics, but also by the variation
of atmospheric conditions. Therefore long exposures in a large field
of view require off-axis guiding to be performed using at least two
guide stars. Whereas one star is used to correct the guiding of the
telescope mounting, the second star serves to take into account the
error of the compensator of the rotation of the field of view. The
latter is maximal in observations made in the prime vertical and at
small zenith angles, where the rate of change the parallactic angle
is the fastest. The telescope was initially equipped with a standard
Ricci cassette meant for photographic observations in the prime focus
within a $18\arcmin$-diameter field. This device had two movable
off-axis guiding microscopes.

Mounting of heavy-weight electrooptical devices (long-slit,
multiobject, and integral-field spectrographs, focal reducers, etc.)
in the prime focus of the 6-m telescope for recording images in a
large field of view made it necessary to seek new design solutions
for off-axis guiding. The use of schemes like the Cassegrain adapter
of the 3.6-m ESO telescope~\cite{ESO77} in the prime focus is fraught
with a number of problems. The high focal ratio ($f/4$) and small
offset of the focal plane above the plane of the rotating table make
it impossible to implement the optomechanical scheme with movable
fields using lens collectives.

An adapter was developed for the prime focus of the 6-m telescope. It
underwent a stage-by-stage upgrade, which included incorporating
results of new technologies and software solutions. As of now, there
have been four versions of this adapter:
%\begin{enumerate}
\begin{list}{}{
\setlength\leftmargin{5mm} \setlength\topsep{0mm}
\setlength\parsep{0mm} \setlength\itemsep{2mm} } \item [1.] In the
first version of the adapter guiding field images were transferred to
a single field via optical fiber bundles. The entrance ends of the
bundles could move in the radial and tangential directions in the
focal plane of the telescope, and the exit ends were fixed in the
focal plane of the guiding eyepiece. Stars were searched manually by
the observer, and guiding by stars was performed visually. On the 6-m
telescope such guiding scheme was first used in 1979 for the optical
identification of 5C survey radio sources using McMullan
electronographic camera~\cite{McM77}. The good guiding quality made
it possible to acquire very deep images with this camera and observe
faint galaxies down to 25th magnitude~\cite{ika80}.

\item [2.] The next version of the adapter was equipped with a
wavelength calibration lamp, and the images from fiber bundles were
transferred to LI-702 TV camera and preamplified with an image
converter tube. The adapter was controlled by a remote controller and
a long communication link from the observers' room located inside the
dome of the 6-m telescope. Beginning with 1990 this  adapter was used
for observations with three versions of MPFS integral-field
spectrograph~\cite{mpfs1,mpfs2} and MOFS multiobject
spectrograph~\cite{mofs}. It was also used for the first observations
with SCORPIO multimode prime focus reducer~\cite{Sco1}.

    \item [3.] Beginning with 2001, observations with SCORPIO and MPFS instruments
continued with a new adapter having fundamentally upgraded design:
lens correctors of guiding fields were mounted to correct the coma of
the primary mirror, a ``flatfield'' lamp, a bundle focusing
mechanism, and an adjustable crosshair illumination were added, and
calibration optics were upgraded to ensure telecentric illumination
from the integrating sphere. The movable units and lamps were
controlled by a microprocessor, which, in turn, was controlled by the
host computer located at the prime focus of the telescope. Since 2007
the TV tube in the guide view was  replaced by a \mbox {CCD camera}.
Automatic search for guide stars was implemented based on USNO\,2.0
catalog. This adapter version, which was described
in~\cite{Sco1,Sco1large}, is still used (summer of 2017) with SCORPIO
instrument for remote observations performed from the Main Laboratory
building of the SAO RAS, 5~km from the 6-m telescope.
%\end{enumerate}
\end{list}

In this paper we present the new, fourth version of the adapter. It
was designed based on our wide experience from previous
modifications, and incorporating such capabilities as correction of
the telescope focus during exposures and control of atmospheric
transmittance. Section~2 describes the overall layout of the adapter
and particularities of its implementation; Section~3 describes
guiding and the properties of the view, Section~4 discusses the
features of the calibrating unit, and the Conclusions section
presents the main points of the work.

\section{DESCRIPTION OF THE ADAPTER}
\subsection{General Scheme}

The adapter is mounted on a rotating table in the prime focus of the
6-m telescope and is used both for offset guiding and for telecentric
illumination of the entrance pupil of the instrument installed on it
by various light sources. The adapter can be used for mounting
SCORPIO-2 focal reducer and other spectrographs with mass no greater
than 150~kg and flange focal distance no greater than 40~mm.

\begin{figure*}[]
    \setcaptionmargin{5mm}
    \onelinecaptionsfalse
    \centerline{\includegraphics[scale=0.27]{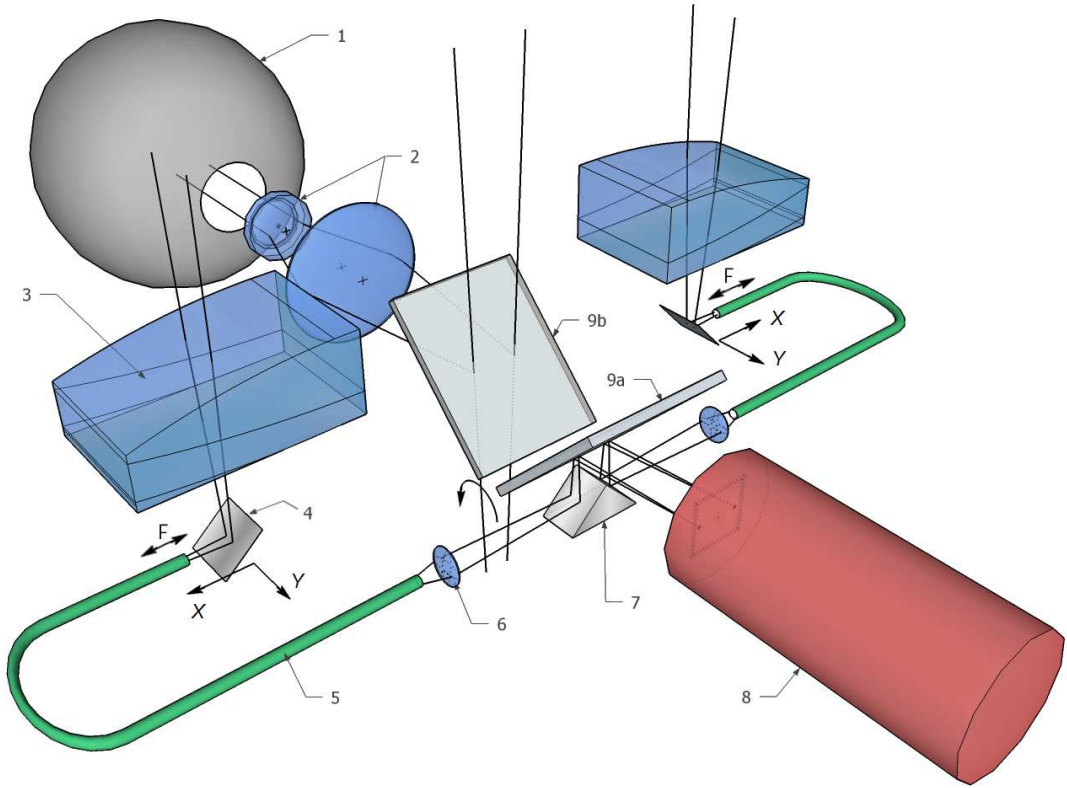}}
    \caption{Optical layout of the adapter:
    (1)~--- integrating sphere,
    (2)~--- calibration illuminator optics,
    (3)~--- off-axis lens corrector,
    (4)~--- mirror,
    (5)~--- fiber bundle,
    (6)~--- fiber lens,
    (7)~--- reflecting prism,
    (8)~--- guiding CCD,
    (9)~--- diagonal mirror
    (both positions are shown:
    (9a)~--- {FIBERS},
    (9b)~--- {FIELD}).}
    \label{adapter}
\end{figure*}

Figure~\ref{adapter} shows the optical layout of the adapter. The
adapter has two movable fields of view to search for guide
(reference) stars. The light from such a star first comes to off-axis
lens corrector~(3), which compensates coma of the primary mirror of
the telescope. Then, diagonal mirror~(4)redirects the light beam to
the entrance end of fiber bundle~(5), which together with the mirror
is moved across the field in two perpendicular directions by STANDA
linear translation stages\footnote{\texttt{http://www.standa.lt/}}.
The fiber bundle has a diameter of $54\arcsec$ in sky projection, and
can move within a \mbox{$10'\times4\farcs5$} field. The centers of
guide fields are located at a distance of $12\arcmin$  from the
center of the fields of view of the system. The relative arrangement
of the fields is presented in Subsection~\ref{sec_stars}. The images
of the exit ends of the bundles are projected by lens~(6) and
directed by prism~(7) and diagonal mirror~(9) onto the focal plane of
guiding CCD~(8). Mirror~(9) has two working positions~--- FIBERS~(9a)
and FIELD~(9b). Figure~\ref{adapter} shows both positions. In the
FIBERS position light from observed objects is received by the
equipment mounted on the adapter  (the mirror is offset from the
beam), and the guiding CCD captures images of guide stars. In the
FIELD position the diagonal mirror transfers the image of the science
field of view to the guide view (the mirror covers the central beam),
making it possible to perform field identification and to position
bright objects onto the slit. The size of the  FIELD field on the
view is \mbox {$3\arcmin\times2\arcmin$}. The same mirror position is
used to calibrate the main CCD using integrating sphere~(1). For a
more detailed description of the calibrating unit see Section~4.

\subsection{Off-axis Lens Corrector}

\begin{figure}[]
    \setcaptionmargin{5mm}
    \onelinecaptionsfalse
    \centerline{\includegraphics[scale=0.425]{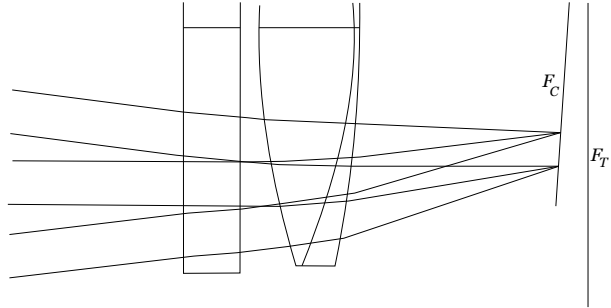}}
    \caption{Optical layout of the off-axis lens corrector.
    }
    \label{offset_layout}
\end{figure}

The main problem in offset guiding on the 6-m telescope is coma
aberration of the parabolic mirror. Standard prime focus corrector
compensates aberrations in the  $18\arcmin$ large field and cannot be
used in our scheme, because the instruments mounted onto the adapter
work without lens corrector. We developed and constructed a
relatively simple two-lens corrector made of domestic-manufactured K8
and TFZ glass. The corrector is achromatized in the 0.5--0.7~$\mu$m
wavelength interval and yields an equivalent telescope focal distance
of 18.5~m. Figure~\ref{offset_layout} shows the optical layout of the
corrector. The focal surface of the corrector $F_C$ within the range
of motion of the fiber bundle is flat to the first approximation. It
is tilted by $3\fdg9$  with respect to the telescope focal
plane~$F_T$.

\begin{figure}[]
    \setcaptionmargin{5mm}
    \onelinecaptionsfalse
    \centerline{\includegraphics[scale=0.3]{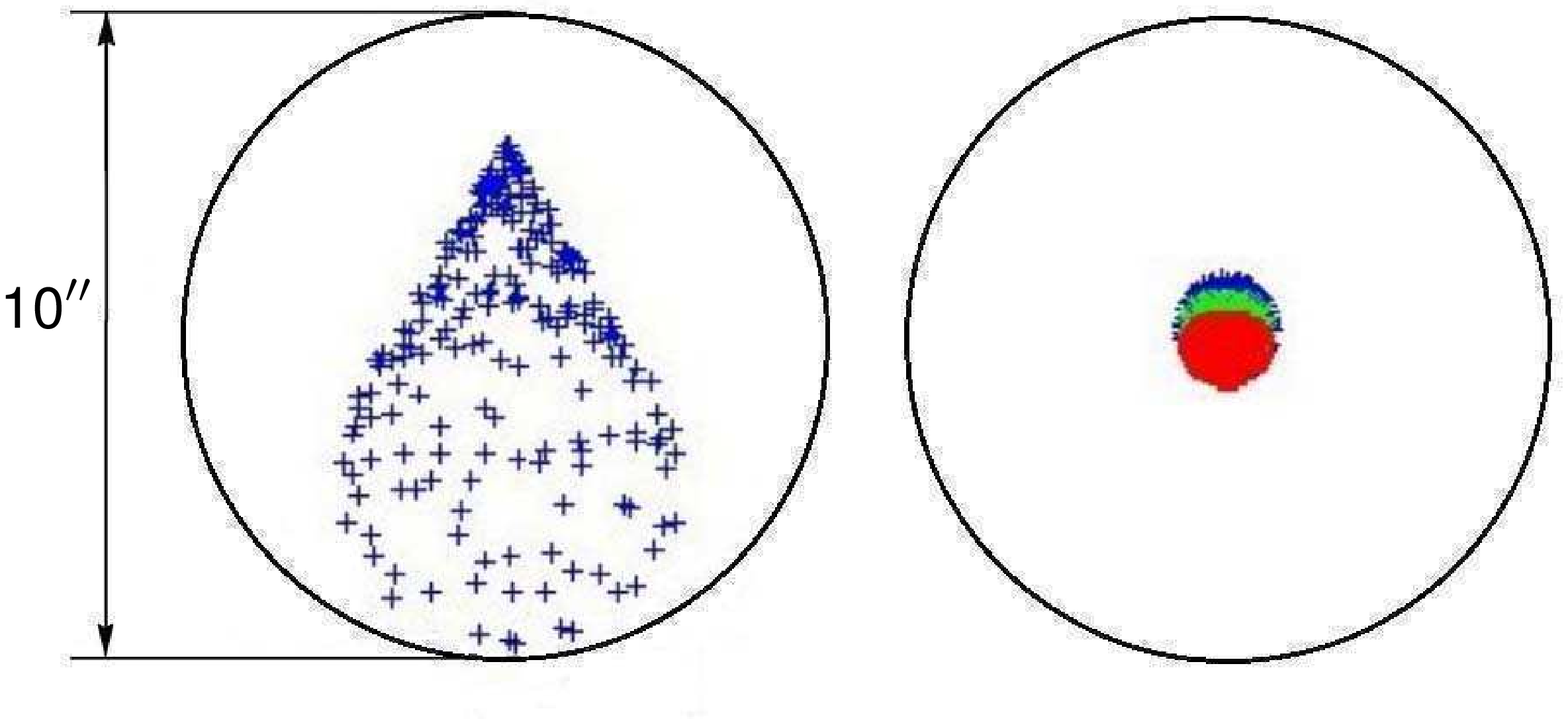}}
    \caption{Computed point diagrams at about $12\arcmin$ offset from the
axis of the primary mirror: without  (left) and with  (right) the corrector.}
    \label{before_after}
\end{figure}

\begin{figure}[]
    \setcaptionmargin{5mm}
    \onelinecaptionsfalse
    \centerline{\includegraphics[scale=0.58]{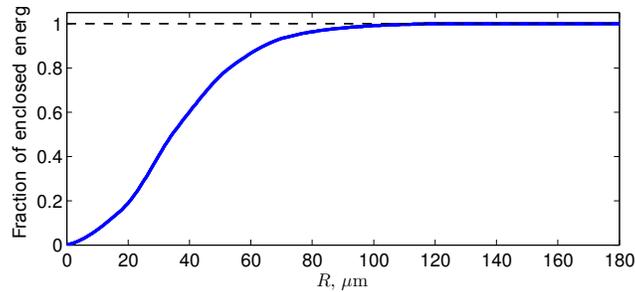}}
    \caption{Mean energy concentration in the scattering circle (spot) in the
$8$--$15\arcmin$ interval of offsets from the optical axis.
    }
    \label{psf}
\end{figure}

Figure~\ref{before_after} shows the result of coma correction at an
offset of about $12\arcmin$ from the primary mirror axis.
Figure~\ref{psf} shows the computed polychromatic point-spread
function of the corrected image. As is evident from the figure, the
seeing ($FWHM$) of the image produced with the use of the corrector
is 40~$\mu$m or better, which corresponds to about $1\arcsec$; it is
quite sufficient for guiding purposes. The optical layout was
computed using ZEMAX
program\footnote{\texttt{http://www.zemax.com/}}.

\subsection{Guide View}

We first used \mbox {LI-702} TV transmitting tubes with a
preamplifier based on an image converter tube with a microchannel
plate as guide views. Since 2007 standard industrial 1/2-inch
monochrome Sony~ICX 429ALL CCD sensor with \mbox {$582 \times 752$}
pixels is used as the guide view. Its parameters are controlled by
RS-232 interface, and the image is digitized by a TV tuner. By the
end of 2017 it is planned to be replaced by commercial Atik
Titan\footnote{\texttt{https://www.atik-cameras.com/product/atik-titan/}}
camera based on a similar  Sony~ICX424 sensor with the images
transmitted to the computer via USB interface.

\begin{figure*}[]
    \setcaptionmargin{5mm}
    \onelinecaptionsfalse
    \centerline{\includegraphics[scale=0.17]{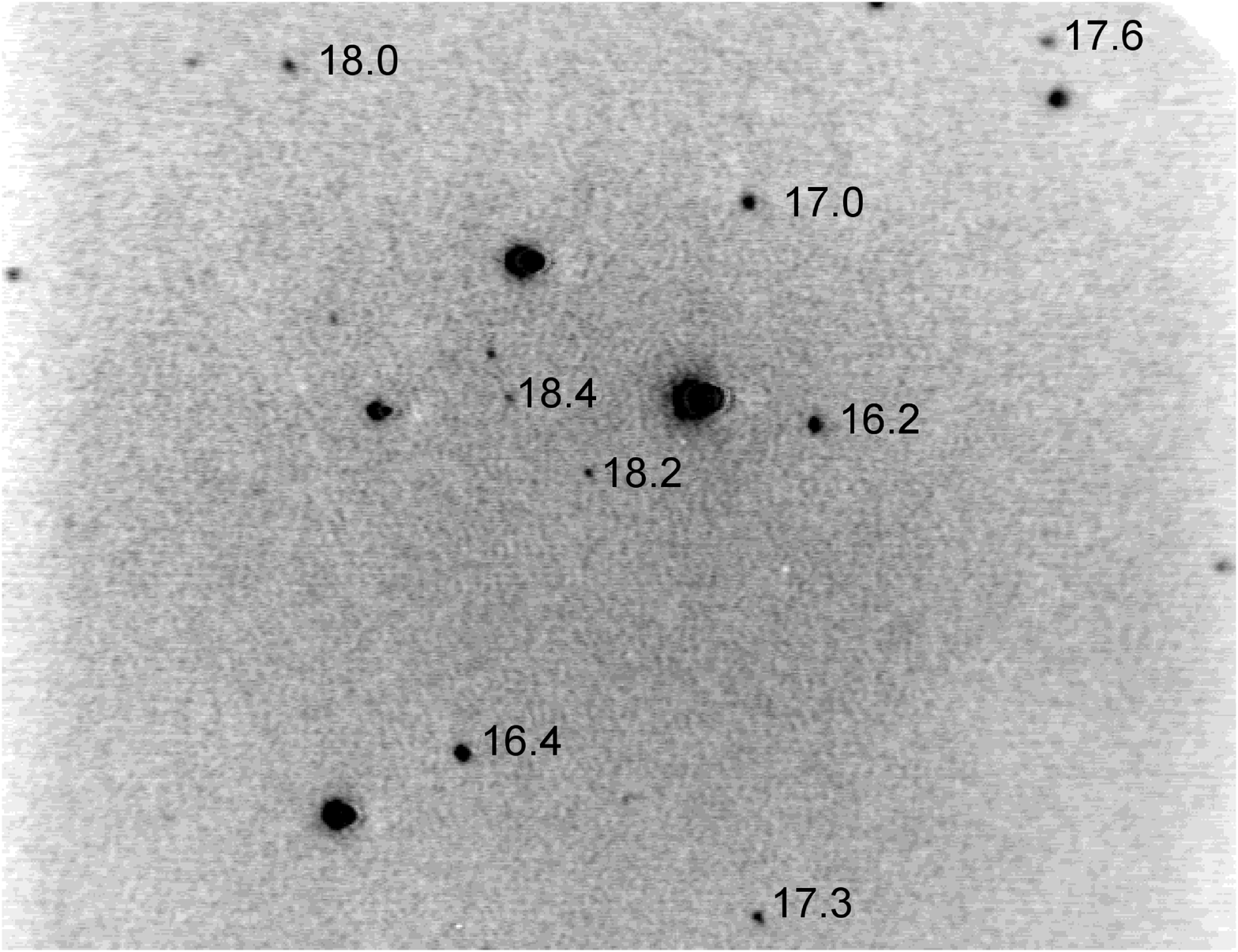}
                \includegraphics[scale=1.20]{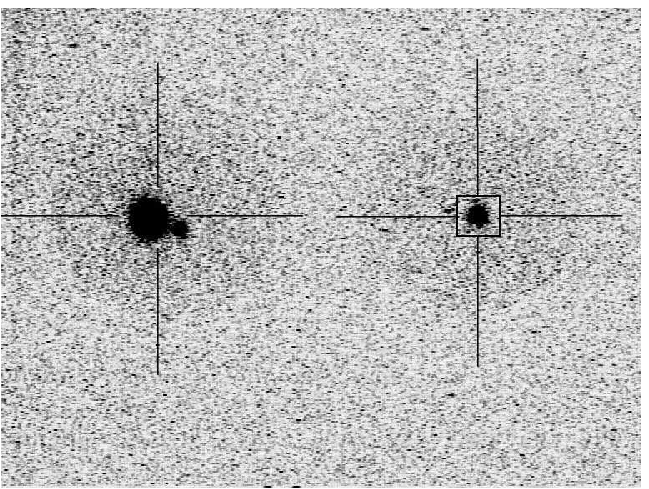}}
    \caption{Example of images acquired in the FIELD (left) and FIBERS (right)
modes with USNO\,2.0 $R$-band magnitudes of faint stars shown.
    }
    \label{tv_field}
\end{figure*}

Figure~\ref{tv_field} (left) shows an example of an image of the main
science field---FIELD. For clarity of illustration we also show the
$R$-band magnitudes of some stars. In the FIELD mode the size of the
field of view is $3\arcmin\times2\arcmin$. The right panel in
Fig.~\ref{tv_field} shows an example of an image from fibers (FIBERS
mode). Because of absorption in fibers, transmission in the FIBERS is
twice lower than in the FIELD mode. However, the sensitivity of the
guiding CCD is sufficient for reliable detection of moonless sky
background.

\subsection{Control System}
\label{sec_remote}

The adapter is controlled like other  \mbox {SCORPIO-2} units.
Two boards have been developed that are equipped with ATmega8535L
microprocessors and power chips to control seven step motors and
adapter calibrations. One board controls four motors used to move two
guide fields along the $X$ and $Y$ axes and two motors for
focusing these fields. Another board controls the FIELD/FIBERS diagonal
mirror switch, line spectrum lamp (NEON), continuum spectrum lamp (QUARTZ),
and  32 LEDs that produce the flatfield.

The ranges of all movements are controlled by limit switches. The current position
$(X, Y)$ of guide fields and the focus are computed from the number of steps made
from the initial position. External control commands are transferred to the
microprocessor via standard RS-422 interface from the industrial computer operating under
 Windows OS. Microprocessor programs are set for autonomous operation:
after receiving a command they run it and report the result when
requested by the control computer. The adapter control interface is
written in IDL
language\footnote{\verb"http://www.harrisgeospatial.com/"} and is
incorporated into the overall  SCORPIO-2 control software.

\section{GUIDING}

After pointing the telescope to  the target object and before
starting the guiding the observer should perform a number of
preliminary procedures including a search for guide stars, selection
of optimum brightness on the view, and preliminary centering. Then
the guiding application is run and if the position of the star image
centroid differs from the position set by the observer in the guide
view the program sends the appropriate correction commands to the
telescope control program.

\subsection{Program for Searching for Guide Stars}

\begin{figure*}[]
    \setcaptionmargin{5mm}
    \onelinecaptionsfalse
    \centerline{\includegraphics[scale=1.03]{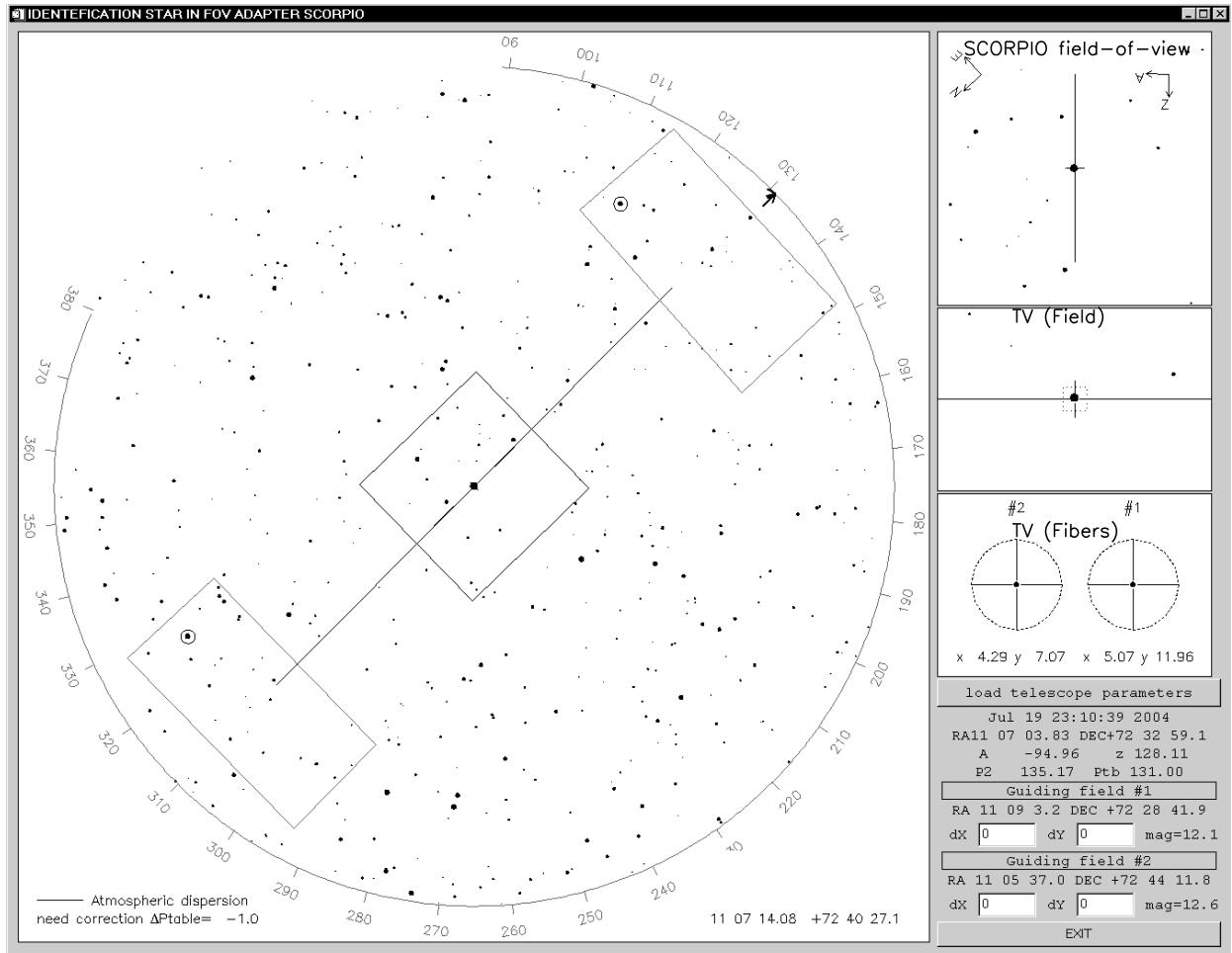}}
    \caption{Program interface for searching for guide stars.}
    \label{identstar}
\end{figure*}

To accelerate pointing and the search for guide stars, we use
IDENTSTAR program written in  IDL, which allows computing the current
instrumental coordinates of guide stars in the adapter field of view
based on the coordinates and orientation of the rotating table
transmitted from the control server of the 6-m telescope.
Figure~\ref{identstar} shows the appearance of the program interface
for searching for guide stars.

When  started, the program reads the current telescope coordinates
from the server of the 6-m telescope and selects from  USNO\,2.0
catalog the coordinates of all stars brighter than $20^{\mathrm m}$
located in the given sky area within the radius of $15\arcmin$. The
large central square in the window marks the position of the
SCORPIO-2 field of view ($6\arcmin\times6\arcmin$), and the
rectangles indicate the limits of the motion of guide fields inside
which the program searches stars for guiding. The small circles whose
sizes correspond to the diameters of fiber bundles, indicate the
brightest stars, which are automatically selected by the  program.
Their magnitudes and positions in the coordinate system of the
adapter are indicated in the bottom right fields of the program
window. If necessary, the observer can select another guide star
among those available in the field. The computed star configuration
in terms of FIELD/FIBERS positions is displayed together with the
star chart from the catalog. Observational experience shows that it
usually takes no more than several minutes to search for and set the
guide stars. Note that preliminary setting of stars can be performed
while the telescope is repositioned from one object to another.

\subsection{Guiding Program}
\label{sec_stars}

For automatic guiding based on a selected star we use our TVGuide
program written in IDL, whereas the C++ program written by
E.~A.~Ivanov for operating AverMedia type TV tuners is used to
digitize the view image at a 25~Hz frequency. In addition to
visualising the guide view image the program can also superimpose
electronic crosshairs and labels. Depending on the adapter mirror
position in the large window, the program also displays either the
slit position in the FIELD field with the target object or two
crosshairs used to aim stars in the FIBERS guiding fields.

\subsection{Limiting Magnitudes}

In the guiding process the tuner polls the guide view output at a
frequency of 25~frames per second; to  improve the noise properties
the images are moving averaged with a 2~to~50 frame window.

\begin{figure}[]
%   \hspace{0.5cm}
    \setcaptionmargin{5mm} \onelinecaptionstrue \captionstyle{normal}
    \includegraphics[scale=0.6]{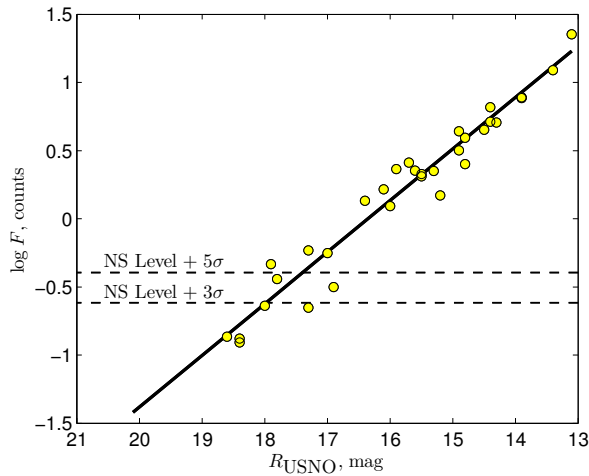}    %    scale
    \caption{Flux logarithms  $\log F$ of guide stars vs. their
${R_{\rm USNO}}$-band magnitudes. Images were captured in the
10-frame moving average mode.}
    \label{limit}
\end{figure}

\begin{figure*}
        \setcaptionmargin{5mm} \onelinecaptionstrue \captionstyle{normal}
%\hspace{-1cm}
\includegraphics[width=0.78\textwidth,trim={0 3.5cm 0 3cm},clip]{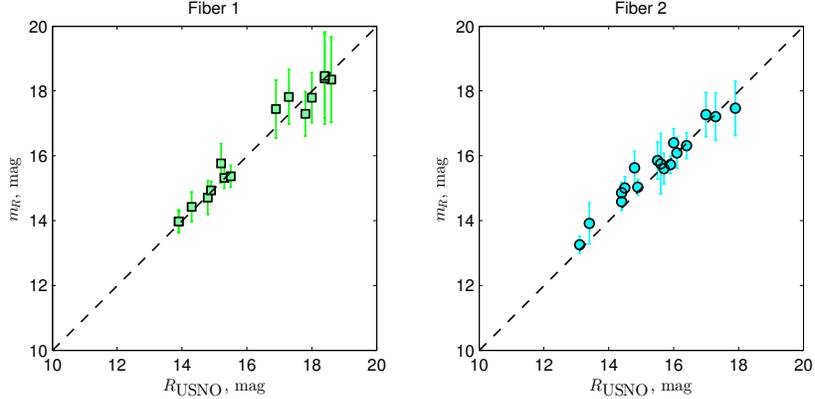}
\caption{ Calibration of the guide view:
instrumental magnitudes $m_{R}$ (ordinates) determined
with the computed zero point plotted against USNO\,2.0 $R$-band magnitudes (abscissae). } \label{tv_calib}
\end{figure*}

The limiting magnitudes of stars suitable for guiding were measured
in November 2015 under clear sky and \mbox
{$\theta=1\arcsec-1\farcs5$} seeing conditions. Figure~\ref{limit}
plots the flux estimates for guide stars against their USNO\,2.0
magnitudes. Intensities were precorrected to take into account the
gain of the CCD view, which is set separately for each guide star.
The dashed lines show the night-sky noise levels for the
$3\sigma_{\rm NS}$ and $5\sigma_{\rm NS}$ detection thresholds. The
regression line crosses the dashed lines at $18\fm0$ and $17\fm4$,
yielding the $R$-band limiting magnitudes for the signal-to-noise
ratios of \mbox {$S/N = 3$} and $5$, respectively. The optimum
magnitude range for guiding is  $10^{\rm m}$--$15^{\rm m}$. Guiding
by stars fainter than $15^{\rm m}$
is possible only under conditions of good seeing and sufficient transmittance. %,  , ,   .

Note that because of the peculiarities of the control system of the 6-m telescope
automatic coordinate correction is performed with a frequency no higher than 0.1~Hz.

\subsection{Control of Atmospheric Transmittance and Guiding Quality}

The zero point for each  fiber bundle of the guide view was
calibrated using observations of real stars under good transmittance
conditions. The results of this calibration are shown in
Fig.~\ref{tv_calib}.

After the calibration procedures were performed it became possible to
control atmospheric transmittance using guide stars. Note that the
calibration relations are further refined during observations.
According to our estimates, systematic errors of atmospheric
absorption measurements due to the difference between the response
curves of the view detector and the $R_{\rm USNO}$-band filter do not
exceed  $0\fm5$.

\begin{figure*}[]
\setcaptionmargin{5mm} \onelinecaptionstrue \captionstyle{normal}
\hspace{0cm}
\includegraphics[width=0.8\textwidth]{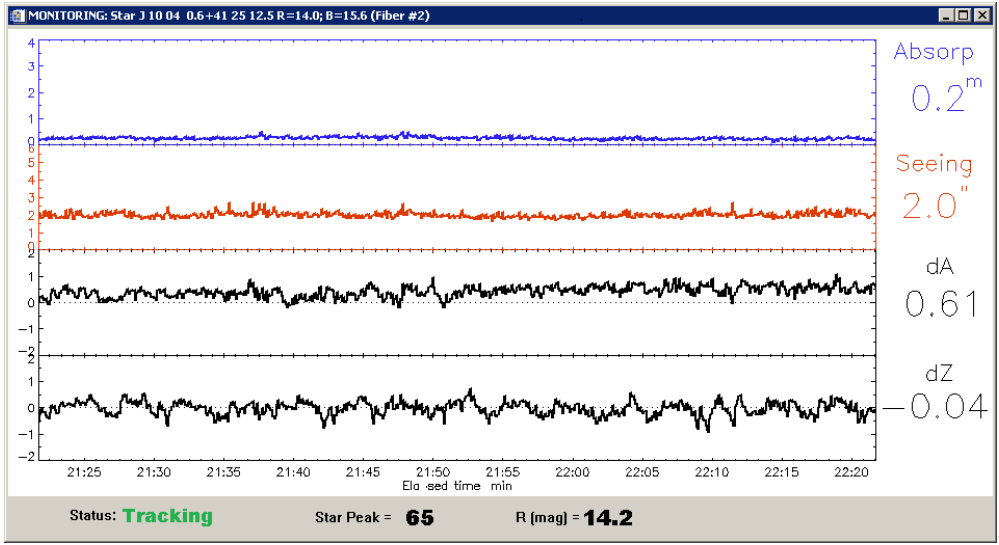}
\caption{Interface of the monitoring window of the image of the guide
star. The window title gives the coordinates of the star, USNO\,2.0
$R$- and $B$-band magnitudes, and the number of the fiber bundle.
    The plots (from top to bottom): $R$-band absorption, seeing, offset of the
centroid of the guide star along the azimuth and zenith distance (in
arcsec). Right: the current parameter values. Bottom: the status
line.
    }
    \label{tv_monitoring}
\end{figure*}

To control the guiding quality, the user may display the plot showing
the current absorption, the size of the image of the guide star and
the azimuth and zenith distance offsets (see
Fig.~\ref{tv_monitoring}).

\subsection{Focusing}

The entrance end of each guiding fiber bundle can move along the
optical axis within 0--9.7~mm to achieve focusing of the guide star
(see Fig.~\ref{adapter}). This makes it possible to correct the
telescope focus during a long  (more
 one hour) series of spectroscopic observations. Observational experience on the
6-m telescope shows that during that time variations of the focal
length caused by thermal deformations of the telescope tube beams may
exceed 0.1--0.2~mm, which is quite significant in the case of good
seeing \mbox{($\theta <2\arcsec$)}. With the telescope and guiding
microscopes focused before the beginning of the exposure series, the
best focus position can then be controlled using the mechanism of
focusing of guiding fields, and a compensating correction can be
applied to adjust the focal  distance of the telescope. The presence
of two guiding fields allows the focus to be controlled by one of the
stars while the other field is used for telescope tracking.

\section{CALIBRATION UNIT}

\subsection{General Description}

The calibration unit consists of an integrating sphere (Ulbricht
sphere), calibration illuminator optics, and control system. The
integrating sphere has two line-spectrum illuminators, a
continuum-spectrum illuminator, and 32 ports for mounting
illuminating LEDs. The line-spectrum source used is a
He--Ne--Ar-filled SGZ-3S gas-discharge radio lamp. The sources of
continuum spectrum are an Osram Halopar-16 50W quartz tube and an
array of 32 LEDs.

The optics of the calibration illuminator produces at the reducer input a
converging beam with the focal ratio equivalent to the focal  ratio of the
telescope ($f/4$). The image of the uniformly illuminated area at the input of the
integrating sphere is formed at the same place where the image of the telescope
mirror is located (the output pupil) and hence the telecentrism condition is fulfilled.
The projected diameter of the illuminator area coincides with the diameter of the
pupil. In our case the telecentrism condition is fulfilled to within
0.2\% or better. Such a solution makes it possible to perform correct
wavelength calibration using line-spectrum lamp and calibrate the transmission
variations of the
system across the field in different modes of spectroscopic observations
(``flatfield'').

The output area of the integrating sphere is illuminated in three ways:
\begin{list}{}{
\setlength\leftmargin{2mm} \setlength\topsep{2mm}
\setlength\parsep{0mm} \setlength\itemsep{2mm} }
    \item [$\bullet$] {NEON}: by a He--Ne--Ar-filled lamp for wavelength calibration.
It is used in the modes of slit and integral-field spectroscopy, and also in
observations with the scanning Fabry--Perot interferometer~\cite{ifp}.
Calibration is usually performed every time when the zenith distance of the
telescope changes by more than $10$--$20\degr$ with the allowance for
instrument flexures.

    \item [$\bullet$] {QUARTZ}: by a halogen-filled quartz-tube con\-ti\-nu\-um-spectrum lamp
for creating ``flatfield.'' Calibration is usually performed at the
beginning and at the end of observations made in the scanning
interferometer mode, and several times during the night in
spectroscopic observations made at different zenith distances. A
serious downside of such illuminator is the abrupt decrease of the
lamp brightness in the blue spectral range, because the intensity
maximum is at the wavelength of about 1200~nm. As a result, at
wavelengths $\lambda<500$~nm the observed spectrum of the lamp is
significantly contaminated by the scattered light from the
longer-wavelength part of the spectrum. The use of equalizing filters
slightly improves the situation.

\item [$\bullet$] {LEDS}: by a system of light-emitting diodes that
forms continuum spectrum for the  ``flatfield'' with approximately
uniform intensity distribution over a wide range of wavelengths. This
allows performing uniformly precise flatfield illumination in
different spectral ranges and reducing stray light in the blue
region. Such a solution was first proposed for photometric
calibration on small telescopes~\cite{sed05,sed13} and it has so far
not been used on large telescopes. For example, a system of 18 LEDs
is planned to be used in the  ``white light'' illuminator of  EXPRES
spectrograph~\cite{expres}.
\end{list}

\begin{figure*}[]
        \setcaptionmargin{5mm} \onelinecaptionstrue \captionstyle{normal}
    \hspace{-1cm}
    \includegraphics[width=0.7\textwidth]{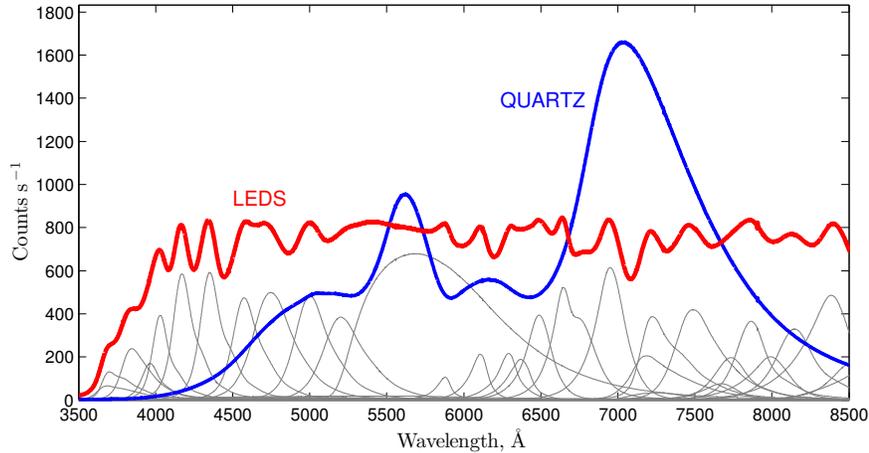}
    \caption{Spectral ``flatfields'' for VPHG940@600 grism
    obtained with  QUARTZ quartz tube (with an equalizing filter) and an array
of light-emitting diodes (LEDS).
    The numbers indicate the spectral curves of the corresponding LEDs.
    Spectra are normalized to 1-s exposure.}
    \label{940_600}
\end{figure*}

\begin{figure*}[]
        \setcaptionmargin{5mm} \onelinecaptionstrue \captionstyle{normal}
    \hspace{0cm}
    \includegraphics[width=0.8\textwidth]{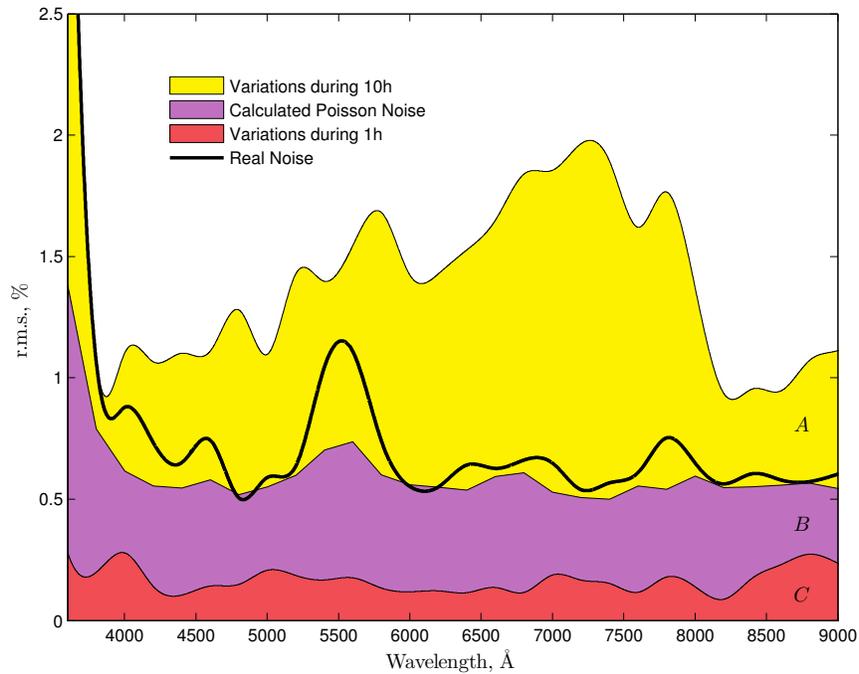}
    \caption{
    Wavelength dependence of relative variations of spectral ``flatfield'' with
    VPHG\,940@600 grism obtained with LED illumination.
    }
    \label{normstab}
\end{figure*}

Let us now consider in detail the latter illumination option provided by light-emitting
diode system LEDS.

\subsection{Light-Emitting Diode System}

Modern industry offers a wide range of various light-emitting diodes
(LEDs) for the entire optical wavelength range. Each LED emits
continuum spectrum in a narrow, about 10--20~nm wide wavelength
interval. The brightness of individual elements in such a LED array
can be selected so as to  construct spectral ``flatfield'' for
various grisms taking into account spectral sensitivity of the
instrument mounted on the adapter.

We use pulse-width modulation to control the brightness of individual LEDs.
To each LED a rectangular-wave signal is sent with different duty ratio, which
varies discretely from 0 to 253. Such realization is easy to implement and it
ensures high stability of LED brightness. LEDs are controlled by the microprocessor
mounted on the second board (see Section~\ref{sec_remote}). The  program started at
the control computer in the  IDL environment allows choosing either standard
illumination configuration or setting the brightness of LEDs individually.

To illustrate LEDS technique, we show in Fig.~\ref{940_600} an
example of the ``flatfield'' constructed for\linebreak VPHG\,940@600
grism with the operating wavelength interval 3500--8500~\AA.

\subsection{Stability of LEDs}

The calibrating lamp consisting of a set of LEDs must ensure
repeatability of the form of the spectrum in the temperature interval
\mbox {$\Delta T = -20$...$+40\degr$C}. We estimate the stability of
the luminous flux of the diode due to the change of external
temperature  $T$ from the following empirical relation adopted
from~\cite{sch06}:
\begin{equation}
    F=F_{300} \exp\left(- \dfrac{ T[{\rm K}] - 300}{T_1}\right),
    \label{eq1}
\end{equation}
where $F_{300}$ is the LED flux at the close-to-room temperature and
$T_1$ is the characteristic temperature for the given class of LEDs.
The experimental  $T_1$ values for blue, green, and red LEDs -- 1600,
295, and 95~K, respectively, are adopted from~\cite{sch06}.  As is
evident from equation~(\ref{eq1}), the flux of red LEDs is most
sensitive to temperature variations and therefore the variations of
the form of the spectrum should be the strongest in this wavelength
range. Let us estimate its stability adopting the sensitivity of the
flux of red LEDs as the worst case. Let us differentiate expression
(\ref{eq1}):
\begin{equation}
    dF/F_{300}= -\exp\left(-\dfrac{T - 300}{T_1} \right) \dfrac{dT}{T_1}
    \label{eq2}
\end{equation}
and compute the temperature coefficient of flux in the case of the
change of the temperature by one degree.
% ~\ref{templed}

To illustrate the effect, we list in the table the characteristic
values of derivatives in the operating temperature range for red
LEDs.

\begin{table}[]
\setcaptionmargin{0mm} \onelinecaptionstrue
\captionstyle{nonumber}
    \caption{\centerline{Temperature coefficient for red LEDs}}
%\begin{center}
        \begin{tabular}{c|c|c|c|c }
            \hline
             $T, \degr$C & $-20$    & $0$      & $27$      & $40$ \\
            \hline
            $dF/F_{300}$ & $-0.017$ & $-0.014$ & $-0.0105$ & $-0.0009$ \\
            \hline
        \end{tabular}
    %\end{center}
    \label{templed}
\end{table}

When estimating the possible fluctuations of the spectrum of the lamp
one must take into account the fact that during spectral exposure (5
to 60~minutes) the temperature at the primary focus of the telescope
does not change by more than $1$--$2\degr$C.

To estimate the actual stability of LED illumination, two series of
spectral ``flatfields'' were obtained in the laboratory with
SCORPIO-2 using VPHG\,940@600 grism: one series was acquired over 10
hours at 1-hour intervals, and another series was acquired over 1
hour at 1-minute intervals. Figure~\ref{normstab} shows the relative
variations of the ``flatfield'' as a function of wavelength. Domain
$A$ shows the scatter for the 10-hour experiment, which approximately
simulates the situation for a single observing night. For comparison,
we show domain $B$, which corresponds to computed Poisson noise. We
calculated Poisson noise as $1/\sqrt{N_{s} \times gain}$, where
$N_{s}$ is the number of counts at the given wavelength and $gain$ is
the gain of the CCD. The thick line shows the actual noise in the
spectrum, which agrees well with the theoretical level. Domain $C$
shows the scatter for the 1-hour experiment, which approximately
simulates the situation for observations of a single object. One-hour
variations ($C$) can be seen to be lower than Poisson noise ($B$). We
determined the variations of both data series and the Poisson noise
within 200~\AA-wide bins.

Note that despite small variations of the ambient temperature over 10
hours (they remain within $5\degr$C limits), the amplitude of
variations can be seen to gradually increase with wavelength from 1\%
to 2\%. Variations drop to 1\% after 7500~\AA\ and this behavior
reflects the contribution of the second order of VPHG\,940@600 grism,
and illumination in this wavelength range is also produced by blue
LEDs.

Thus the ``flatfield'' produced using LEDS technique yields more uniform illumination
compared to the ``flatfield'' produced using QUARTZ technique
(Fig.~\ref{940_600}). It would be reasonable to suggest that the hybrid
QUARTZ+LEDS ``flatfield'', where  LEDS corrects the drawbacks of QUARTZ technique
in the blue part of the spectral range, should be the optimum solution.
However, the LEDs and the quartz tube  are located too close to each other, and the quartz
tube heats up strongly when turned on, and therefore the LED properties change when
both types of illumination are used simultaneously. This fact should be taken into  account
in observations and when designing similar systems. In the case of spectroscopic
observations performed during the night we recommend using  LEDS before QUARTZ
(if both types of illumination are necessary), and separate their use in
calibrations before and after observations.

The fact that LEDs are now widely used and relatively inexpensive suggests that such
technique can be easily implemented on modern spectrographs.

\section{CONCLUSIONS}

The latest, fourth version of  the adapter, which, like the previous
two versions, was developed at the Laboratory of spectroscopy and
photometry of extragalactic objects of the Special Astrophysical
Observatory of the Russian Academy of Sciences, has been used for
observations with SCORPIO-2 focal reducer~\cite{Sco2} since 2012 .

Compared to the previous version of the adapter, which is still
(summer 2017) used with  SCORPIO, the new version has substantially
reduced chromatic aberrations in guiding fields, and the spectral
flatfield calibration is now formed by the combined emission of the
LED array, which made it possible to increase the signal in the blue
part of the spectrum multiple times.

Let us now point out the main features of the presented adapter:
%\begin{enumerate}
\begin{list}{}{
\setlength\leftmargin{5mm} \setlength\topsep{2mm}
\setlength\parsep{0mm} \setlength\itemsep{2mm} }
 \item [1.] To transfer the images of the guide stars to the guiding CCD two fiber
bundles are used with possible individual focusing during observations.

\item [2.] The use of off-axis lens corrector made it possible to
correct coma of the guide star images up to about $15\arcmin$ from
the optical axis.

\item[3.] The optics of the calibration illumination is made to fulfil the telecentrism
condition.

\item [4.] LED system allows achieving relatively uniform  spectral illumination
of the CCD taking into account its response curve and the transmission curves of the
grisms.

\item[5.] Photometry of the images of guide stars during observations
is used to qualitatively control atmospheric transmittance and
focusing of the telescope.
\end{list}
%\end{enumerate}

The guiding and calibration system described has been successfully tested
for many years on the 6-m telescope of the Special Astrophysical Observatory
of the Russian Academy of Sciences.

%     {acknowledgements}

\begin{acknowledgments}
 This work is supported by the Russian Science Foundation (grant no.~17-12-101335).
We are grateful to E.~A.~Ivanov for developing the control program for
the TV tuner.
\end{acknowledgments}

%\bibliographystyle{AstroBull}
%\bibliography{Biblio}  % bibtex database
%\end{document}

\onecolumngrid

\begin{flushright}
{\it Translated by A.~Dambis}
\end{flushright}

\end{document}